\documentstyle[12pt,epsf]{article}

\newlength{\pubnumber} 

\catcode`\@=11
\@addtoreset{equation}{section}

\def\section{\@startsection{section}{1}{\z@}{3.5ex plus 1ex minus .2ex}
 {2.3ex plus .2ex}{\large\bf}}
\def\subsection{\@startsection{subsection}{2}{\z@}{2.3ex plus .2ex}
 {2.3ex plus .2ex}{\bf}}

\def\beq{\begin{equation}}
\def\eeq{\end{equation}}
\def\beqn{\begin{eqnarray}}
\def\eeqn{\end{eqnarray}}

\def\nolabel{\nonumber }

\def\mod#1{{\rm \,\, (mod\, #1)}}

\def\half{{\textstyle{1\over 2}}}
\def\third{{\textstyle {1\over3}}}
\def\fourth{{\textstyle {1\over4}}}

\def\twothird{{\textstyle {2\over3}}}

\def\Tr{{\rm Tr}\, }

\def\bone{{\mathbf 1}}
\def\mone{{\mathbf 1}}

\def\mo{{\mathbf 0}}
\def\mzero{{\mathbf 0}}
\def\mF{{\mathbf F}}
\def\mQ{{\mathbf Q}}
\def\mS{{\mathbf S}}

\def\ma{{\mathbf a}}
\def\mb{{\mathbf b}}
\def\mk{{\mathbf k}}
\def\mV{{\mathbf V}}
\def\bV{{\mathbf V}}

\def\mbeta{ {{\mathbf \beta}}}

\def\eps{\epsilon}

\def\H#1{H_{#1}}

\def\FD2pv{FD2$^{'}$V }
\def\FD2p{FD2$^{'}$ }

\def\inbar{\,\vrule height1.5ex width.4pt depth0pt}

\def\IC{\relax\hbox{$\inbar\kern-.3em{\rm C}$}}
\def\IQ{\relax\hbox{$\inbar\kern-.3em{\rm Q}$}}
\def\IR{\relax{\rm I\kern-.18em R}}
 \font\cmss=cmss10 \font\cmsss=cmss10 at 7pt

\def\IZ{\relax\ifmmode\mathchoice
 {\hbox{\cmss Z\kern-.4em Z}}{\hbox{\cmss Z\kern-.4em Z}}
 {\lower.9pt\hbox{\cmsss Z\kern-.4em Z}}
 {\lower1.2pt\hbox{\cmsss Z\kern-.4em Z}}\else{\cmss Z\kern-.4em Z}\fi}

\def\IZT{\IZ_2\times\IZ_2}

\def\Io{\relax\ifmmode\mathchoice
 {\hbox{\cmss 1\kern-.4em 1}}{\hbox{\cmss 1\kern-.4em 1}}
 {\lower.9pt\hbox{\cmsss 1\kern-.4em 1}}
 {\lower1.2pt\hbox{\cmsss 1\kern-.4em 1}}\else{\cmss 1\kern-.4em 1}\fi}

\def\u{\underline{\phantom{a}}}

\begin{document}

\begin{titlepage}
\samepage{
\setcounter{page}{1}
\rightline{BU-HEPP-05/03}
\rightline{CASPER-05/03}
\rightline{\tt hep-ph/0506183}
\rightline{June 2005}
\vfill
\begin{center}
 {\Large \bf  Observable/Hidden Sector Broken Symmetry\\
              for Symmetric Boundary Conditions\\}
\vfill
\vfill {\large
        B.~Dundee,\footnote{Ben$\u$Dundee@baylor.edu} 
        J.~Perkins,\footnote{John$\u$Perkins@baylor.edu} and 
        G.~Cleaver,\footnote{Gerald$\u$Cleaver@baylor.edu}
}
\\
\vspace{.12in}
{\it        Center for Astrophysics, Space Physics \& Engineering Research,\\
            Department\ of Physics, Baylor Science Building, Baylor University,\\
            Waco, TX 76706, USA\\}
\vspace{.025in}
\end{center}
\vfill
\begin{abstract}
A 4-dimensional heterotic string model of free fermionic construction is presented 
wherein mirror symmetry breaking between observable and hidden sector gauge groups 
occurs in spite of mirror symmetry between observable and hidden sector 
worldsheet fermion boundary conditions. 
The differentiation is invoked 
by an asymmetry in GSO projections necessarily resulting from the symmetry of the 
free fermionic boundary conditions.
In the specific examples shown, an expected non-chiral 
Pati-Salam mirror universe model is transformed into a chiral model with enhanced 
hidden sector gauge symmetry and reduced observable sector gauge symmetry: 
$[SU(4)_C\otimes SU(2)_L\otimes SU(2)_R]^{O} \otimes 
 [SU(4)_C\otimes SU(2)_L\otimes SU(2)_R]^{H}$,  
 is necessarily transformed into a chiral
$[SU(4)_C\otimes SU(2)_L]^{O}\otimes 
 [SO(10)\otimes SU(2)_R]^{H}$ model because of an unavoidable
 asymmetry in GSO projections.
\end{abstract}
\smallskip}
\end{titlepage}

\setcounter{footnote}{0}

\section{Mirror Universe Models}

Mirror universe models, with hidden sector matter and forces identical to that 
of the visible world, but interacting with the latter only via gravity, have been 
proposed in the context of neutrino physics (specifically by attempts to 
understand the nature of a sterile neutrino) and in the context of 
superstring/M-theory \cite{mu1,mu2,mu3,mu4,mu5}. As a means
to reconcile the constraints of big bang nucleosynthesis, the reheating
temperature of the mirror universe after inflation was postulated to be lower than that in the observable 
universe \cite{mu1}. From this it was shown that the asymmetric reheating can be related to a 
difference of the electroweak symmetry breaking scales in the two sectors, as required for a 
mirror solution to the neutrino puzzle. In such models it was shown that the baryon asymmetry
is greater in the mirror universe than in the observable universe and that the mirror baryons could provide 
the dominant dark matter in the bulk universe \cite{mu4}.

While heterotic strings (M-theory) offer a mirror universe scenario in 
ten (eleven) dimensions, with the well-known $E_8$ gauge symmetry for both observable and hidden 
sectors (9-branes), string-derived quasi-realistic three generation mirror universe models in 
four-dimensions have, to our knowledge, not been constructed. 
Rather, in typical quasi-realistic string models, the ten-dimensional mirror symmetry is 
broken through compactification to four dimensions.
In bosonic lattice/orbifold construction differentiation of observable 
and hidden gauge groups (and, thus, of matter representations) is typically a result of 
asymmetric Wilson loop effects and spin embedding. 
In corresponding free fermionic \cite{fff} models, 
mirror symmetry breaking generally results from 
asymmetric boundary conditions between observable and hidden sector worldsheet fermions.
An investigation into string/M-derived mirror models is underway and will be reported  
in an upcoming paper \cite{DPC2005b}. 
Contrastingly, in this letter we wish to discuss
an obstacle to mirror model construction that can develop in weakly coupled heterotic strings. 
This obstacle, nonetheless, produces an interesting physical effect in its own right. 
In this letter we will demonstrate how, under a specific set of conditions,
mirror symmetry breaking will necessarily occur in models with symmetric
boundary conditions. We will investigate the implications of this for initially mirror symmetric
models.

During our string/M-derived mirror universe investigation,
we found that even when symmetric worldsheet fermion boundary conditions are imposed,
breaking of mirror symmetry is sometimes mandated by an unavoidable asymmetry imposed by 
the GSO projections.  
In this note we show, through two example models, how 
GSO projections can necessitate mirror symmetry breaking of observable ($O$) and hidden sector ($H$)
Pati-Salam \cite{psm1,psm2} gauge groups,
\beqn
[SU(4)_C\otimes SU(2)_L\otimes SU(2)_R]^{O}\otimes 
[SU(4)_C\otimes SU(2)_L\otimes SU(2)_R]^{H}\, .
\label{psmgg}
\eeqn
In our first example, the GSO projections reduce the observable sector gauge group to 
\beqn
[SU(4)_C\otimes SU(2)_{L}]^{O},
\label{obsbr}
\eeqn 
by transferring $SU(2)_{R}^{O}$ 
to the hidden sector. In the process the 
\beqn
[SU(4)_C\otimes SU(2)_L]^{H} 
\label{hsen1a}
\eeqn
subgroup of the hidden sector Pati-Salam gauge 
symmetry is enhanced to
\beqn
[SO(10)]^{H}\, .
\label{hsen1b}
\eeqn
Additionally, the initial shadow $(S)$ sector (corresponding to charges possibly 
carried by both observable and hidden states)
gauge group, $[SU(2)^3\otimes SU(3) \otimes U(1)^7]^{S}$, absorbs 
$[SU(2)_{R}]^{H}$ to become
$[SU(2)^3\otimes SU(5)\otimes U(1)^6]^S$ (see Table B.2).

Our second example differs from the first in some of the elements of its GSO projections matrix. 
For model 2,
the GSO projections reduce the observable sector Pati-Salam gauge group to 
$[SU(4)_C\otimes SU(2)_{R}]^{O}$
by alternately transferring $SU(2)_{L}^{O}$ 
to the hidden sector. The hidden sector $[SU(4)_C\otimes SU(2)_R]^{H}$ subgroup of the Pati-Salam group is 
similarly enhanced to $[SO(10)]^{H}$. In this version, the shadow sector gauge group remains of rank 12 
and does not absorb $[SU(2)_L]^{H}$ (see Table C.2).
 
In Section 2 we briefly review free fermionic construction and the related 
form of the GSO projections.
In Section 3 we then review the NAHE-based $\IZT$ models \cite{nahe}, 
the class to which our example models belong (see Table A.1).  
Following this, in Section 4 we introduce the mirror-symmetric free fermion 
worldsheet boundary conditions for our two models (see Tables A.2 and A.3).
We then demonstrate how an unavoidable asymmetry that arises between certain
GSO projection matrix elements 
in our models produces mirror symmetry breaking between observable and hidden sector states.
We study viable hypercharge definitions in Section 5. Our investigation 
of the unavoidable observable/hidden sector mirror-breaking effect in our models is then
summarized in Section 6.
We present the gauge group and matter states for our first model in Tables B.2 and B.3, 
respectively, and those for our second model in Tables C.2 and C.3, respectively.

\section{Free Fermionic Models}

A free fermionic heterotic string model is specified by
two objects \cite{fff}. The first is a $p$-dimensional basis set 
of free fermionic boundary vectors $\{\mV_i$, $i= 1$, ... , $p \}$. 
Each vector $\mV_i$ has 64 components, $-1< V_i^{m}\leq 1$, $m=1$, ... , 64,
with the first 20 components specifying boundary conditions for the 20 real free fermions 
representing  worldsheet degrees of freedom for the left-moving supersymmetric string, and the
later 44 components specifying boundary conditions for the 44 real free fermions representing 
worldsheet degrees of freedom for the right-moving bosonic string.
(Components of $\mV_i$ for complex fermions are double-counted.) 
Modular invariance dictates that the basis vectors, $\mV_i$, span a finite
additive group $\Xi = \{\sum_{n_i=1}^{N_i -1} \sum_{i=1}^{p}{{n_i}{\mV_i}} \}$,
with $N_i$ the lowest positive integer such that $N_i \mV_i = \mzero \mod{2}$. 
$\mV_1 = \mone$ (a 64-component unit vector)  
must be present in all models.   
In a given sector 
{\boldmath$\alpha$\unboldmath}$\equiv a_i \mV_i \mod{2}\in \Xi$, 
with $a_i\in\{0,\, 1,\, \dots\, , N_i -1\}$, 
a worldsheet
fermion $f_m$ transforms as $f_m\rightarrow -\exp\{ \pi \alpha_m \} f_m$ around 
non-contractible loops on the worldsheet. Boundary vector components for real fermions 
are thus limited to be either 0 or 1, 
whereas boundary vector components for complex fermions can be rational \cite{fff}.

The second object necessary to define a free fermionic model 
(up to vacuum expectation values (VEVs) of fields in the effective field theory)
is a $p\times p$-dimensional matrix $\mk$ of rational numbers 
$-1<k_{i,j}\leq 1$, $i,j= 1$, ..., $p$, that determine the GSO operators 
for physical states. The $k_{i,j}$ are related to the phase 
weights $C{{\mV_i}\choose {\mV_j}}$ appearing in a model's one-loop partition 
function $Z$: 
\beqn
C{{\mV_i}\choose {\mV_j}}
= (-1)^{s_i+s_j}
        {\rm exp}( \pi i k_{j,i}- \half \mV_i \cdot\mV_j), 
\label{kijpw}
\eeqn
where $s_i$ is the 4-dimensional spacetime component of $\mV_i$.
(The inner product of boundary (or charge) vectors is lorentizian, 
taken as left-movers minus right movers. Contributions to inner products 
from real fermion boundary components are weighted by a factor of $\half$ compared to 
contributions from complex fermion boundary components.) 

The phase weights $C{ { 
\mbox{\boldmath$\alpha$\unboldmath}
}
\choose {
\mbox{\boldmath$\beta$\unboldmath}
} 
}$ for general sectors 
\beqn
\mbox{\boldmath$\alpha$\unboldmath}
= \sum_{j=1}^{p} a_j \mV_j \in \Xi ,\,\,\,\,
\mbox{\boldmath$\beta$\unboldmath}
  = \sum_{i=1}^{p} b_i \mV_i \in \Xi 
\label{bab} 
\eeqn
can be expressed in terms of the components in the
$p\times p$-dimensional matrix $\mk$ for the basis vectors:
\beqn
C{{
\mbox{\boldmath$\alpha$\unboldmath}
}\choose {
\mbox{\boldmath$\beta$\unboldmath}
}}
= (-1)^{s_{\alpha}+s_{\beta}}
        {\rm exp}\{ \pi i \sum_{i,j} b_i( k_{i,j} - \half \mV_i \cdot\mV_j) a_j \}.  
\label{kab}
\eeqn

Modular invariance simultaneously imposes constraints on the basis vectors $\mV_i$ 
and on components of the GSO projection matrix $\mk$:
\beqn
k_{i,j} + k_{j,i} &=& \half\, \mV_i\cdot \mV_j\, \mod{2}
\label{kseta1}\\
N_j k_{i,j} &=& 0 \, \mod{2}   
\label{kseta2}\\
k_{i,i} + k_{i,1}&=& - s_i + \fourth\, {\mV_i}\cdot {\mV_i}\, \mod{2}.
\label{kseta3}
\eeqn

The dependence upon the $k_{i,j}$
can be removed from equations (\ref{kseta1}-\ref{kseta3}),
after appropriate integer multiplication, 
to yield three constraints on the $\mV_i$:
\beqn
&&N_{i,j} \mV_i\cdot \mV_j =  0 \, \mod{4} \label{ksetb1}\\
&&N_{i}   \mV_i\cdot \mV_i =  0 \, \mod{8} \label{ksetb2}\\
&&{\rm The~number~of~real~fermions~simultaneously~periodic}\nolabel\\
&&{\rm for~any~three~basis~vectors~is~even.}\label{ksetb3}
\eeqn
$N_{i,j}$ is the lowest common multiple of $N_i$ and $N_j$. 
(\ref{ksetb3}) applies when two or more of three the basis vectors are identical.
Thus, each basis vector must have an even number of real periodic fermions.

The physical massless states in the Hilbert space of a given sector
{\boldmath$\alpha$\unboldmath}$\in{\Xi}$, are obtained by acting on the vacuum with
bosonic and fermionic operators and by
applying generalized GSO projections. The $U(1)$
charges for the Cartan generators of the unbroken gauge group 
are in one to one correspondence with the $U(1)$
currents ${f^*_m}f_m$ for each complex fermion $f_m$, and are given by:
\beq
{Q^{
\mbox{\boldmath$\alpha$\unboldmath}
}_m = {1\over 2}\alpha_m + F^{
\mbox{\boldmath$\alpha$\unboldmath}
}_m},
\label{u1charges}
\end{equation}
where $\alpha_m$ is the boundary condition of the worldsheet fermion $f_m$
in the sector {\boldmath$\alpha$\unboldmath}, and
$F^{
\mbox{\boldmath$\alpha$\unboldmath}
}_m$ is a fermion number operator counting each mode of
$f_m$ once and of $f_m^{*}$ minus once. Pseudo-charges for non-chiral (i.e., with 
both left- and right-moving components) real Ising fermions $f_m$ can be similarly 
defined, with $F_m$ counting each real mode $f$ once.
 
For periodic fermions,
$\alpha_m =1$, the vacuum is a spinor representation of the Clifford
algebra of the corresponding zero modes.
For each periodic complex fermion $f_m$
there are two degenerate vacua ${\vert +\rangle},{\vert -\rangle}$ ,
annihilated by the zero modes $(f_m)_0$ and
$(f_m^{*})_0$ and with fermion numbers 
$F^{
\mbox{\boldmath$\alpha$\unboldmath}
}_m =0,-1$, respectively.

The boundary basis vectors $\mV_j$
generate the set of GSO projection operators for physical states from all sectors
{\boldmath$\alpha$\unboldmath}$ \in \Xi$.
In a given sector {\boldmath$\alpha$\unboldmath}, the surviving states are those that
satisfy the GSO equations imposed by all $\bV_j$ and determined by the 
$k_{j,i}$'s:
\beqn
 \mV_j\cdot {\mF}^{
\mbox{\boldmath$\alpha$\unboldmath}
} = \left(\sum_i k_{j,i} a_i\right) + s_j
    - \half\, \mV_j\cdot {
\mbox{\boldmath$\alpha$\unboldmath}
}\, \mod{2} ,
\label{gso1-a}
\eeqn
or, equivalently,
\beqn
\bV_j\cdot {\mQ}^{
\mbox{\boldmath$\alpha$\unboldmath}
} = \left(\sum_i k_{j,i} a_i\right) + s_j\, \mod{2}.
\label{gso}
\eeqn
For a given set of basis vectors, the independent GSO matrix components
are $k_{1,1}$ and $k_{i,j}$, for $i>j$.    
This GSO projection constraint, when combined with equations (\ref{kseta1}-\ref{kseta3})
form the free fermionic re-expression of the even, self-dual  
modular invariance constraints for bosonic lattice models.

\section{$\IZT$ NAHE-Based Models}

The $\IZT$ NAHE-based models have contributed much to understanding 
regions of the parameter space of weakly-coupled (quasi)-realistic string models, including
GUT's \cite{fsu5}, semi-GUT's \cite{psm2,sguts}, near-MSSM's \cite{nmssm,optun}, and MSSM's \cite{mssm}.
They have advanced knowledge of general features of classes of string model, both perturbative \cite{ffgenprop}
and non-perturbative properties \cite{ffnp} that likely still persist after M-theory embedding.
The NAHE set \cite{nahe} consists of the five basis vectors, $\bone$,
$\mS$, and $\mb_i$, $i=1$ to 3, given in Table A.1. 
The NAHE set breaks 
the observable sector $E_8$ gauge symmetry to $SO(10)\otimes U(1)^3$ 
but does not affect the hidden sector $E_8$. 
The six compactified directions, represented by six pairs 
of left- and right-moving real fermions
$\{ y,\omega \vert \bar{y},\bar{\omega} \}^{1,\cdots,6}$,
generate an additional $SO(12)$ shadow sector (S) gauge symmetry. The
NAHE set breaks this to $SO(4)^3$. However, each $U(1)_i$ combines with a 
corresponding $SO(4)_i$ to form an enhanced $SO(6)_i$ symmetry. 
Thus, the gauge group of the NAHE set is 
$[SO(10) \otimes SO(6)^3]^O \otimes E_8^{H}$ with $N=1$ spacetime supersymmetry.
The observable sector matter is 48 spinorial $\mathbf 16$'s of $SO(10)$, with sixteen of these 
from each sector $\mb_1$, $\mb_2$ and $\mb_3$. 

The $U(1)_i$, for $i=1$ to 3, act as generation charges for the three 
sectors $\mb_1$, $\mb_2$ and $\mb_3$, respectively.
The sixteen copies of $\mathbf 16$'s of $SO(10)$ from a given 
$\mb_i$ are formed from 4 copies of $\mathbf 4$'s of $SO(6)_i$. 
To obtain three generation models, the $SO(6)^3$ 
must be broken to no more than the initial $U(1)^3$.  
Reduction of the $SO(10)$ and of the $SO(6)^3$ can result from  
additional boundary basis vectors that will
break $SO(10)$ to one of its subgroups, $SU(5)\times U(1)$,
$SO(6)\times SO(4)$ or $SU(3)\times SU(2)\times U(1)^2$.
The additional sectors must simultaneously reduce the number of generations to
three, one from each of the sectors $\mb_1$, $\mb_2$ and $\mb_3$.
The detailed phenomenological properties of the various 
NAHE-based three generation models can differ significantly. 
In particular, the properties strongly depend on the boundary components
of the non-NAHE basis vectors  
for the fermions representing the six compactified directions \cite{compactyw}.

\section{Symmetry Breaking of Mirror Models}

Mirror models with matching observable and hidden sector symmetries and states 
may be created from NAHE-based models by adding mirror basis vectors,
$\mb_1^{'}$, $\mb_2^{'}$ and $\mb_3^{'}$, as defined in Table A.2. Due to the symmetry between
$\mb_{i}$ and $\mb_{i}^{'}$, these mirror vectors 
break the hidden sector $E_8$ in the same manner as the NAHE 
set breaks the observable sector $E_8$ into $SO(10)\otimes U(1)^3$. 
Each $\mb_{i}^{'}$ produces 16 copies of {\bf 16}'s of the hidden sector $SO(10)$.  

The right-moving components of the 
$\mb_i$ and the $\mb_i^{'}$ basis vectors that are simultaneously non-zero form 
a subset of $\{y,\omega \vert \bar{y},\bar{\omega}\}^{1,\cdots,6}$. 
That is, $\mb_i$ and $\mb_i^{'}$ states will both carry some
$\{y,\omega \vert \bar{y}, \bar{\omega}\}^{1,\cdots,6}$ ``shadow sector'' 
charges.
Within the NAHE set the only two bosonic sectors that can produce gauge states
are $\mo$ and $\mone + \mb_1 + \mb_2 + \mb_3$. 
All observable and ``shadow sector'' 
$SO(n)$ and $U(1)$ generators, along with the hidden sector $\mathbf 120$ rep of $SO(16)\in E_8$,
originate in $\mo$, 
while the hidden sector $\mathbf 128$ rep of $SO(16)\in E_8$ originates in 
$\mone + \mb_1 + \mb_2 + \mb_3$. 
The GSO projections from the mirror sectors $\mb_1'$, $\mb_2'$, and $\mb_3'$ remove
the $\mathbf 128$ rep of $SO(16)$, reducing the hidden sector $E_8$ symmetry to $SO(16)$.  
Further, the mirror sectors reduce the hidden sector $SO(16)$ to an $SO(10)\otimes U(1)^3$ symmetry,
matching the observable sector.  
(Similarly the GSO projections of $\mb_1$, $\mb_2$, and $\mb_3$ remove
any contribution to the observable gauge group from    
the $\mone + \mb_1' + \mb_2' + \mb_3'$ sector.)

The hidden sector $U(1)^3$ generation charges and the 
observable sector $U(1)^3$ generation charges 
combine with the shadow sector charges in a like manner.  
Simultaneously, the $\mb_i'$ sector GSO's reduce  
the $SO(4)^3$ shadow sector symmetry to $U(1)^6$. The net result is a 
$SU(3)\otimes SU(2)^3 \otimes U(1)^7$ shadow sector symmetry, whose charges are carried by 
matter representations of the observable $SO(10)$ and of the hidden $SO(10)$.  
The additional $SU(3)\otimes SU(2)^3$ generators originate in
several additional massless gauge sectors formed from linear combinations of 
$\mone$, $\mb_i$, and $\mb'_i$. All of these sectors have massless vacua that
only carry $\bar{\eta}_i$, $\bar{\eta}_i'$, and $\bar{y}$, $\bar{\omega}$ shadow charges.      
While most massless gauge states from these sectors are projected out, 
a few are not. Those that survive combine 
observable sector $\bar{\eta}$-charges,
hidden sector $\bar{\eta}'$-charges,
and shadow sector $\bar{y},\bar{\omega}$-charges (see Tables B.2 and C.2).
Thus, at this stage mirror symmetry still exists, as should be expected.    
Note in particular that $\mk$ is invariant under exchange of $\mb_i$ with co responding $\mb'_i$.

To break each of the $SO(10)$ to their corresponding Pati-Salem 
$SU(4)_C\otimes SU(2)_L\otimes SU(2)_R$, two additional sectors 
$\ma$ and $\ma'$ are added to the model, where $\ma$ and $\ma'$ are mirror sectors, as shown in   
in Table A.2. The set of 10 basis vectors $\{\mone,\mS,\mb_1,\mb_2,\mb_3,\mb_1',\mb_2',\mb_3',\ma,\ma'\}$,
produce a model with mirror symmetry boundary conditions for the observable and hidden sectors. 
However, with the addition of $\ma$ and $\ma'$, an {\it unavoidable} asymmetry develops among the GSO 
projection components:
Although all new degrees of freedom of 
$k_{
\mbox{\boldmath$\alpha$\unboldmath}
,
\mbox{\boldmath$\beta$\unboldmath}
}$, 
for 
{\boldmath$\alpha$\unboldmath}$\in \{ \ma, \ma' \}$ and 
{\boldmath$\beta$\unboldmath}$\in \{ \mone,\mS,\mb_1,\mb_2,\mb_3,\mb_1',\mb_2',\mb_3' \}$,
are chosen to be invariant under simultaneous exchange of primed and un-primed sectors for both
{\boldmath$\alpha$\unboldmath} and 
{\boldmath$\beta$\unboldmath}, 
symmetry breaking between $k_{\ma,\ma'}$ and $k_{\ma',\ma}$ occurs automatically.  
Only one of $k_{\ma,\ma'}$ and $k_{\ma',\ma}$ is a degree of freedom; the other is 
specified by (\ref{kseta1}). As Table A.3 shows, $\ma\cdot \ma' = 10$, which from (\ref{kseta1}) yields
\beqn
k_{\ma,\ma'} + k_{\ma',\ma} &= 1 \, \mod{2}.
\label{ksetc1}
\eeqn  
Thus, $k_{\ma,\ma'}$ and $k_{\ma',\ma}$ cannot be equal (mod 2);   
either $k_{\ma,\ma'}= 1$ and $k_{\ma',\ma}= 0$ or vice versa 
(since all components of $\ma$ and $\ma'$ are either anti-periodic or periodic). 

The addition of $\ma$ and $\ma'$ to the model generates several new massless gauge sectors
of the form $\ma + \ma' + ...\, $. 
However, in both model variations presented herein, the GSO projections remove all possible 
gauge states from such sectors,
except those coming from $\ma + \ma'$. The $\ma + \ma'$ sector has all anti-periodic
components except for four periodic associated with the two complex fermions generating 
the observable $SO(4) = SU(2)_L \otimes SU(2)_R $ and the two complex 
fermions generating the hidden sector $SO(4) = SU(2)_L \otimes SU(2)_R $.  
Thus, in the $\ma + \ma'$ sector, massless gauge states require one anti-periodic fermionic 
(with $Q= \pm1$) excitation.  

The GSO projection from $\ma$ acts on observable $SO(4)$ spinors while $\ma'$ acts 
on hidden $SO(4)$ spinors. Since $k_{\ma,\ma'}$ and $k_{\ma,\ma'}$ differ by 1 (mod 2),
so do $k_{\ma,\ma+\ma'}$ and $k_{\ma',\ma+\ma'}$. Thus,
a state in the $\ma + \ma'$ sector survives both the $\ma$ sector GSO projections and
$\ma'$ sector GSO projections if and only if its observable and hidden $SO(4)$ spinors  
have opposite chirality. That is, the net number of down spins among the four spinors must be odd, 
implying that an $\ma + \ma'$ gauge state will either carry 
observable $SU(2)_L\in SO(4)$ charge and hidden $SU(2)_R\in SO(4)$ charge or
vice-versa. 

For model 1, the
additional $k_{\mb^{(')}_i,\ma}$ and $k_{\mb^{(')}_i,\ma'}$ require $\ma+\ma'$ states to always
have 
an even number of observable $SO(4)$ $-\half$ spins and 
an odd  number of hidden     $SO(4)$ $-\half$ spins, 
linking observable $SU(2)_R$ with hidden $SU(2)_L$. 
The remaining GSO projections on $\ma+\ma'$ gauge states require the $Q= \pm 1$ anti-periodic 
fermion excitation charge to be from the hidden sector $SO(6)\sim SU(4)$.
Thus, the surviving $\ma + \ma'$ simple root connects 
observable $SU(2)_R$ roots with the hidden sector $SO(6)$ and $SU(2)_L$ roots,
thereby regenerating a hidden sector $SO(10)$ (see Table B.2):
\beqn
[SU(2)_{R}]^O\otimes [SO(6)\otimes SU(2)_{L}]^H\rightarrow [SO(10)]^H\, .
\label{2lenh}
\eeqn
 
In addition, a sector $1 + \mb_1 + \mb_1' + \ma$ gauge state
with eight complex spinors links the 
shadow gauge states with the hidden sector $SU(2)_{R}$, increasing the shadow sector gauge symmetry,
\beqn
[SU(3)\otimes SU(2)^3 \otimes U(1)^7 ]^S \otimes (SU(2)_{R})^H
\rightarrow [SU(5)\otimes SU(2)^3 \otimes U(1)^6]^S\, .
\label{2lenh2}
\eeqn
The final model 1 gauge group is, therefore,
\beqn
[SU(4)\otimes SU(2)_L]^O \otimes [SU(5)\otimes SU(2)^3 \otimes U(1)^6]^S \otimes [SO(10)]^H\, .
\label{gg1}
\eeqn

While maintaining symmetry under exchange of primed and un-primed components of $\mk$,
model 2 differs from model 1 in some choices of    
$k_{\ma,
\beta}$ and $k_{\ma',
\beta}$, for
{\boldmath$\mbeta$\unboldmath}$\in \{ \mone,\mS,\mb_1,\mb_2,\mb_3,\mb_1',\mb_2',\mb_3' \}$
(see Table C.1). 
Model 2 GSO projections require an odd number of observable $SO(4)$ $-\half$ spins and 
an even number of hidden $SO(4)$ $-\half$ spins in the $\ma + \ma'$ sector. 
This link observable $SU(2)_L$ reps with hidden $SU(2)_R$ reps. 
The remaining GSO projections again require the $Q= \pm1$ anti-periodic 
fermion excitation charge of an $\ma + \ma'$ simple root to be from the hidden sector $SO(6)$. 
Thus, for model 2 the gauge boson from $\ma + \ma'$ again regenerates a hidden sector $SO(10)$:
\beqn
[SU(2)_{L}]^O\otimes [SO(6)\otimes SU(2)_{R}]^H\rightarrow [SO(10)]^H\, .
\label{2lenh3}
\eeqn
However, in model 2, no additional gauge state is produced to mix the shadow sector with the hidden
sector $SU(2)_{L}$. 
The final model 2 gauge group is, therefore,
\beqn
[SU(4)\otimes SU(2)_R]^O \otimes [SU(3)\times SU(2)^3 \times U(1)^7]^S \otimes 
[SO(10)\otimes SU(2)_L]^H\, .
\label{gg2}
\eeqn
(An exchange of definitions of left and right-handedness in the observable sector and of related 
SM states transforms this into 
\beqn
[SU(4)\otimes SU(2)_L]^O \otimes [SU(3)\times SU(2)^3 \times U(1)^7]^S \otimes 
[SO(10)\otimes SU(2)_L]^H\, .)
\label{gg3}
\eeqn

\section{Hypercharge Definitions}

An important issue for these models is whether an acceptable definition of hypercharge 
can be found, since the conventional hypercharge is missing either its equivalent $T_3^R$ 
contribution in model 1 or its equivalent $T_3^L$ contribution in model 2. In standard 
NAHE-based models the hypercharge is formed as
\beqn
Y = \third \tilde{Q}_C +  \half \tilde{Q}_L,
\label{hyp1}
\eeqn
(for $Y(Q_L) = \third$ normalization),
where $\tilde{Q}_C = \sum_{m=1}^{3}Q_{\bar{\psi}^m}$ is the associated charge trace of 
$U(1)_C \equiv [\bar{\psi}^{1*}\bar{\psi}^{1} + \bar{\psi}^{2*}\bar{\psi}^2 
                     + \bar{\psi}^{3*}\bar{\psi}^3 ]$ and
$\tilde{Q}_L = \sum_{m=4}^{5}Q_{\bar{\psi}^m}$ is the associated charge trace of 
$U(1)_L \equiv  [\bar{\psi}^{4*}\bar{\psi}^4 + \bar{\psi}^{5*}\bar{\psi}^5 ]$.
Since $SU(4)_C\rightarrow SU(3)_C \otimes U(1)_{B-L}$, 
$\third \tilde{Q}_C = \tilde{Q}_{B-L}$, as Table 4.a indicates. Similarly, since 
$SO(4) = SU(2)_L \otimes SU(2)_R \rightarrow SU(2)_L \otimes U(1)_L$,
$\half \tilde{Q}_L = 2 T_3^R$.
Thus,
\beqn
Y = \tilde{Q}_{B-L} + 2 T_3^R , 
\label{hyp2}
\eeqn       
which yields electromagnetic charge
\beqn
\tilde{Q}_{EM} = T_3^L + \half Y = T_3^L + T_3^R + \half \tilde{Q}_{B-L}.
\eeqn       
Hence, model 1 requires a replacement for $T_3^R$, 
while  model 2 needs a replacement for $T_3^L$.

Under $SU(4)_C \otimes SU(2)_L$ the SM $SU(3)_C \otimes SU(2)_L$ left handed reps combine into
\begin{itemize}
\item $Q_L   = ({\mathbf 3},{\mathbf 2})_{2T_3^R = 0}  
\oplus   L_L = ({\mathbf 1},{\mathbf 2})_{2T_3^R = 0} \rightarrow 
(QL)_L = ({\mathbf 4},{\mathbf 2})_{2T_3^R = 0}$, 
\item $e^{c}_L = ({\mathbf 1},{\mathbf 1})_{2T_3^R = 1} 
\oplus d^{c}_L = ({\mathbf 3},{\mathbf 1})_{2T_3^R = 1}
\rightarrow 
(d^{c} e^{c})_L = ({\mathbf 4},{\mathbf 1})_{2T_3^R = 1}$, 
\item $\nu^c_L  = ({\mathbf 1},{\mathbf 1})_{2T_3^R = -1} 
\oplus u^{c}_L = ({\mathbf 3},{\mathbf 1})_{2T_3^R = -1}\rightarrow 
(u^{c} \nu^{c})_L = ({\mathbf 4},{\mathbf 1})_{2T_3^R = -1}$.
\end{itemize}
Model 1 contains three generations of pairs   
of $(q^{c}l^{c})^{n}_i = (\mathbf{4},\mathbf{1})$ states
(with generation index specified by $i= 1$ to $3$, pair element index specified by  $n= 1, 2$, 
and left-handed index L implicit), 
where $(q^cl^c)$ denotes either $(d^c e^c)$ or $(u^c\nu^c)$ (see Table B.3). 
These states are also respective 
doublets under the three generation $SU(2)_i$ of the shadow sector. As discussed,
for a three generation model,
the $SU(2)_i$ must be broken to the  
generational $U(1)_i$ of standard NAHE models by additional GSO projections from further sectors. 
When each $SU(2)_i$ is broken to $U(1)_i$ in this manner, one  
component of each $SU(2)_i$ doublet is also projected out. If these additional GSO 
projections can be chosen such that the up-spin component of $SU(2)_i$ for $(q^{c}l^{c})^{n=1}_i$ survives 
along with the down-spin component of $SU(2)_i$ for $(q^{c}l^{c})^{n=2}_i$, then 
for $Y = \tilde{Q}_{B-L} + 2 (T_3)_i$
the $(q^{c}l^{c})^{n=1}_i$ become the $(d^{c}e^{c})_i$ states and the $(q^{c}l^{c})^{n=2}_i$ become the 
$(u^{c}\nu^{c})_i$ states. 
Under $L\leftrightarrow R$ exchange, the same process can be applied to create a consistent three generation 
hypercharge for model 2. 

For both models, this is the only possible choice for $(q^{c}l^{c})$ hypercharge, since in each model
the extra Abelian charges carried by the $(q^{c}l^{c})^{n}_i$ are independent of the index $n$ for each 
generation $i$.  
That is, no hypercharge definition involving only the extra $U(1)_k$ could yield valid
hypercharge for both $(d^{c}e^{c})$ and $(u^{c}\nu^{c})$ reps. For model 1, 
this posses a difficulty for the MSSM higgs.
In model 1, the only additional $SU(2)_L$ doublets are singlets under all $SU(2)_i$. 
These are the pairs of 
states $h^n_i$ ($n=1,$ $2$) and the more exotic $H^n_i$, which are also {\bf 5} reps of $SU(5)^S$. 
For a given generation $i$, the extra $U(1)_k$ charges are independent of the index $n$. 
Thus, no hypercharge definition could yield both a $(Y = -1)$-charged 
up-higgs and a $(Y = +1)$-charged down-higgs from an $h^{n=1}_i$ and $h^{n=2}_i$ pair. 
Instead, a hypercharge definition is required such 
that the $U(1)_k$ hypercharge contribution to $QL$ and $ql$ states is zero, 
while it is +1 for at least one $h^{n}_i$ pair and -1 for at least one other $h^{n}_{i'}$ pair.       
Applying the six $QL$ and $(q^{c}l^{c})$ constraints prevents
any $U(1)_k$ from appearing in a general hypercharge definition.  
Thus, model 1 cannot provide a suitable definition for hypercharge, 
unless the Cartan subalgebra of $SU(5)^S$ can
contribute to the hypercharge of $H^n_i$ components after $SU(5)^S$ is broken. 

In contrast, the MSSM higgs of model 2 come in the standard $h_i$ and $\bar{h}_i$ pairs for each 
generation i. However, 
each $h_i$ and $\bar{h}_i$ is also an $SU(2)_i$ doublet (see Table C.3).
Thus, if under $SU(2)_i$ breaking by
GSO projections from additional sectors, the $SU(2)_i$ up-spin component of $h_i$ survives and the
$SU(2)_i$ down-spin component of $\bar{h}_i$ (or vice-versa) then 
$h_i$ becomes down-higgs and $\bar{h}_i$ becomes up-higgs (or vice versa). 
Thus, a viable hypercharge definition for
model 2 is
\beqn
Y = \half \tilde{Q}_{B-L} + \sum_{i=1}^{3} T_3^i.
\label{m2hyp}
\eeqn 

This would produce generational higgs pairs, which is a common occurrence 
in NAHE-based models. This often provides for mass hierarchy between generations 
since the physical higgs usually becomes a weighted (by several orders of magnitude)
linear combination of the generational higgs.
MSSM matter states then couple differently by 
generation to the physical higgs, producing a large mass hierarchy, even when all
mass couplings in the superpotential are third order.

Models 1 and 2 both contain an anomalous $U(1)$ and it is unlikely that additional basis vectors 
would remove the anomaly from either model \cite{anomu1}. 
In fact, additional sectors generally increase the anomaly. 
For model 2 the charge traces of the seven $U(1)_k$ are
\beqn
\Tr \mQ = (0,-144,96,0,0,-192,0)\, .
\label{m2trq}
\eeqn 
(as can be computed from Table C.3).
The anomaly may be rotated into a single $U(1)_A$,
\beqn
U(1)_A = [-3 Q_2 + 2 Q_3 - 4 Q_6]
\label{m2qa}
\eeqn
for which the trace is 1392.  The orthogonal
\beqn
&&U_2' =  2 Q_2 + 3 Q_3\label{q2p}\\
&&U_3' = -3 Q_2 + 2 Q_3 + (13/4) Q_6 \, .\label{q3p}
\label{m2qb}
\eeqn
become non-anomalous (traceless).

Model 2 is another example \cite{sguts} for which non-Abelian fields must necessarily take on VEVs 
to cancel the Fayet-Iliopoulos (FI) term,
\beqn
\eps\equiv\frac{g^2_s M_P^2}{192\pi^2}\Tr Q^{(A)} = \frac{g^2_s M_P^2}{192\pi^2} 1392,
\label{fit}
\eeqn
generated in the $U_A$ $D$-term
by the Green-Schwarz-Dine-Seiberg-Witten anomalous U(1) breaking mechanism
\cite{gsdsw}. 
To see this, first note from Table C.3 
that singlet states $S^n_1$ through $S^n_6$ carry the non-anomalous charge $Q_1 = 3$, 
while the remaining singlets have $Q_1 = 0$. 
Thus, $D$-flatness for $U(1)_1$ cannot be maintained if only singlets receive VEVs
and one or more of the fields $S^i_1$ through $S^i_6$ are among those that do. 
Next, the singlets $S_7$ and $\bar{S}_7$ do not carry anomalous charge 
(only $Q_4$ and $Q_5$ charge), and so cannot help cancel the F-I term.  
The remaining singlets are simply 
$S_8$ and $S_9$, and their vector partners of opposite charges.
$S_8$ carries anomalous charge $Q_A = 4$ 
and non-zero non-anomalous charges $Q_4 = 1$, $Q_5 = 4$, $Q'_2 = 6$, $Q'_3 = -42$, 
while
$S_9$ carries anomalous charge $Q_A = 4$ 
and non-zero non-anomalous charges $Q_4 =-1$, $Q_5 =-4$, $Q'_2 = 6$, $Q'_3 = -42$.
$\bar S_8$ and $\bar S_9$ carry respective opposite charges.
Thus, we see that no combination of $S_{7,8,9}$ and $\bar{S}_{7,8,9}$ VEVs
can simultaneously cancel the anomalous $D_A$-term contribution from the trace of $U_A$ and keep the D-terms for
$Q'_2$ and $Q'_3$ flat. Therefore, some non-Abelian fields must take on VEVs in the 
process of cancelling the Tr$\, Q_A$ contribution to $D_A$ to maintain $D$-flatness.     
Of particular interest is whether $SU(2)_i$-charged fields take on VEVs in the parameter 
space of flat directions. Analysis of flat directions for these models is, 
however, beyond the scope of this letter.

\section{Summary}
We have demonstrated how, under certain conditions, 
mirror symmetry is necessarily broken between the observable and hidden sector gauge groups
of heterotic string models with mirror boundary conditions 
for observable and hidden sector worldsheet fermions.
The observable/hidden sector gauge group mirror breaking occurs because of an unavoidable 
asymmetry in GSO projections. 
This effect can be induced in free fermionic models through 
an observable/hidden sector mirror pair of 
basis vectors, $\ma$ and $\ma'$, with the properties that:
\begin{itemize}
\item Their vector sum yields new, independent gauge sectors 
(possibly after further basis vectors are added) $\ma + \ma' + ...$. 
\item They do not overlap with non-zero components in the 
observable and hidden sectors. 
\item Their inner product is not equal $0 \mod{4}$.
\end{itemize} 
Under these conditions the observable and hidden sector gauge states 
from some $\ma + \ma' + ...$ sector (or sectors) 
will not be mirror images, since the observable gauge states surviving the 
$k_{\ma, \ma + \ma' + ...}$ GSO projections 
will be different from the hidden sector gauge states 
surviving the corresponding $k_{\ma', \ma + \ma' + ...}$ GSO projections.
In the examples shown, starting with mirror Pati-Salam gauge groups
$[SU(4)_C\otimes SU(2)_L\otimes SU(2)_R]^{O} \otimes 
 [SU(4)_C\otimes SU(2)_L\otimes SU(2)_R]^{H}$,  
the observable sector
$SU(2)_{R(L)}$ was transformed to the hidden sector
by this necessary asymmetry of the GSO projections,  
enhancing the hidden sector gauge group to
$[SO(10)\otimes SU(2)_{R(L)}]^{H}$.  
This transference of gauge rank from the observable sector to the hidden sector 
acts favorably for 
coupling strength renormalizations, allowing non-Abelian hidden sector 
coupling strengths to increase faster than observable sector coupling strengths,
with decreasing energy below the string scale,  
leading to the formation of generally advantageous intermediate scale hidden sector condensates.

\section{Acknowledgments}

This work is supported in part by the NASA/Texas Space Grant Consortium.

\newpage
\def\AEF{A.E. Faraggi}
\def\APPB#1#2#3{{\it Acta.\ Phys.\ Polon.\ }\/ {\bf B#1} (#2) #3}
\def\APJ#1#2#3{{\it Astrophys.\ Jour.\ }\/ {\bf#1} (#2) #3}
\def\AP#1#2#3{{\it Ann.\ Phys.}\/ {\bf#1} (#2) #3}
\def\NPB#1#2#3{{\it Nucl.\ Phys.}\/ {\bf B#1} (#2) #3}
\def\NPBPS#1#2#3{{\it Nucl.\ Phys.}\/ {{\bf B} (Proc. Suppl.) {\bf #1}} (#2) 
 #3}
\def\PLB#1#2#3{{\it Phys.\ Lett.}\/ {\bf B#1} (#2) #3}
\def\PRD#1#2#3{{\it Phys.\ Rev.}\/ {\bf D#1} (#2) #3}
\def\PRL#1#2#3{{\it Phys.\ Rev.\ Lett.}\/ {\bf #1} (#2) #3}
\def\PRT#1#2#3{{\it Phys.\ Rep.}\/ {\bf#1} (#2) #3}
\def\PTP#1#2#3{{\it Prog.\ Theo.\ Phys.}\/ {\bf#1} (#2) #3}
\def\MODA#1#2#3{{\it Mod.\ Phys.\ Lett.}\/ {\bf A#1} (#2) #3}
\def\MPLA#1#2#3{{\it Mod.\ Phys.\ Lett.}\/ {\bf A#1} (#2) #3}
\def\IJMP#1#2#3{{\it Int.\ J.\ Mod.\ Phys.}\/ {\bf A#1} (#2) #3}
\def\IJMPA#1#2#3{{\it Int.\ J.\ Mod.\ Phys.}\/ {\bf A#1} (#2) #3}
\def\nuvc#1#2#3{{\it Nuovo Cimento}\/ {\bf #1A} (#2) #3}
\def\RPP#1#2#3{{\it Rept.\ Prog.\ Phys.}\/ {\bf #1} (#2) #3}
\def\etal{{\it et al\/}}
               

\def\bibiteml#1#2{ }
\bibliographystyle{unsrt}

\newpage

\appendix
\def\s{\phantom{-}}

{{\bf Table A.1 THE NAHE SET}

\begin{tabular}{|c|c|ccc|c|ccc|c|ccc|}
\hline
&$\psi^\mu$ & ${\chi^{12}}$ & ${\chi^{34}}$ & ${\chi^{56}}$ &
        $\bar{\psi}^{1,...,5} $ &
        {$\bar{\eta}^1$}&
        {$\bar{\eta}^2$}&
        {$\bar{\eta}^3$}&
        $\bar{\psi'}^{1,...,5} $ &
        {$\bar{\eta'}^1$}&
        {$\bar{\eta'}^2$}&
        {$\bar{\eta'}^3$}\\
\hline
\hline
      {\bf 1} &  1 & 1&1&1 & 1,...,1 & 1 & 1 & 1 & 1,...,1 & 1 & 1 & 1 \\
        $\mS$ &  1 & {1}&{1}&{1} & 0,...,0 & 0 & 0 & 0 & 0,...,0 & 0 & 0 & 0 \\
\hline
  {${\mb}_1$} &  1 & {1}&0&0 & 1,...,1 & {1} & 0 & 0 & 0,...,0 & 0 & 0 & 0 \\
  {${\mb}_2$} &  1 & 0&{1}&0 & 1,...,1 & 0 & {1} & 0 & 0,...,0 & 0 & 0 & 0 \\
  {${\mb}_3$} &  1 & 0&0&{1} & 1,...,1 & 0 & 0 & {1} & 0,...,0 & 0 & 0 & 0 \\
\hline
\end{tabular}}

\vspace{0.2truecm}

{\bf \phantom{NAHE SET b}}

{\begin{tabular}{|c|cc|cc|cc|}
\hline
 &      {$y^{3,...,6}$}  &
        {${\bar y}^{3,...,6}$}  &
        {$y^{1,2},\omega^{5,6}$}  &
        {${\bar y}^{1,2},\bar{\omega}^{5,6}$}  &
        {$\omega^{1,...,4}$}  &
        {$\bar{\omega}^{1,...,4}$}   \\
\hline
\hline
    {\bf 1} & 1,...,1 & 1,...,1 & 1,...,1 & 1,...,1 & 1,...,1 & 1,...,1 \\
   $\mS$    & 0,...,0 & 0,...,0 & 0,...,0 & 0,...,0 & 0,...,0 & 0,...,0 \\
\hline
{${\mb}_1$} & {1,...,1} & {1,...,1} & 0,...,0 & 0,...,0 & 
                                          0,...,0 & 0,...,0 \\
{${\mb}_2$} & 0,...,0 & 0,...,0 & {1,...,1} & {1,...,1} & 
                                          0,...,0 & 0,...,0 \\
{${\mb}_3$} & 0,...,0 & 0,...,0 & 0,...,0 & 0,...,0 & 
                                     {1,...,1} & {1,...,1} \\
\hline
\end{tabular}}
\vspace{0.4truecm}

{{\bf Table A.2 Mirror Set }

\begin{tabular}{|c|c|ccc|c|ccc|c|ccc|}
\hline
  & $\psi^\mu$ & ${\chi^{12}}$ & ${\chi^{34}}$ & ${\chi^{56}}$ &
        $\bar{\psi}^{1,...,5} $ &
        {$\bar{\eta}^1$}&
        {$\bar{\eta}^2$}&
        {$\bar{\eta}^3$}&
        $\bar{\psi'}^{1,...,5} $& 
        {$\bar{\eta'}^1$}&
        {$\bar{\eta'}^2$}&
        {$\bar{\eta'}^3$}
\\
\hline
\hline
  {${\mb'}_1$} &  1 & {1}&0&0 & 0,...,0 & 0 & 0 & 0 & 1,...,1 & {1} & 0 & 0 \\
  {${\mb'}_2$} &  1 & 0&{1}&0 & 0,...,0 & 0 & 0 & 0 & 1,...,1 & 0 & {1} & 0 \\
  {${\mb'}_3$} &  1 & 0&0&{1} & 0,...,0 & 0 & 0 & 0 & 1,...,1 & 0 & 0 & {1} \\
\hline
\end{tabular}}
\vspace{0.2truecm}

{\bf \phantom{MIRROR SET b}}

{\begin{tabular}{|c|cc|cc|cc|}
\hline
&  {$y^{1,...,4}$}  
&  {${\bar y}^{1,...,4}$}  &
        {$y^{5,6},\omega^{1,2}$}  &
        {${\bar y}^{5,6},\bar{\omega}^{1,2}$}  &
        {$\omega^{3,...,6}$}  &
        {$\bar{\omega}^{3,...,6}$}   \\
\hline
\hline
{${\mb'}_1$} & {1,...,1} & {1,...,1} & 0,...,0 & 0,...,0 & 
                                          0,...,0 & 0,...,0 \\
{${\mb'}_2$} & 0,...,0 & 0,...,0 & {1,...,1} & {1,...,1} & 
                                          0,...,0 & 0,...,0 \\
{${\mb'}_3$} & 0,...,0 & 0,...,0 & 0,...,0 & 0,...,0 & 
                                     {1,...,1} & {1,...,1} \\
\hline
\end{tabular}}
\vspace{0.4truecm}

{\bf Table A.3 $SO(10)\otimes SO(10)$ Breaking and Generation Reduction}

{\begin{tabular}{|c|c|ccc|c|ccc|c|c|ccc|}
\hline
 & $\psi^\mu$ & ${\chi^{12}}$ & ${\chi^{34}}$ & ${\chi^{56}}$ &
        $\bar{\psi}^{1,...,5}$&
        {$\bar{\eta}^1$}&
        {$\bar{\eta}^2$}&
        {$\bar{\eta}^3$}&
        {$y,\bar{y},\omega\bar, {\omega}$}&
        $\bar{\psi'}^{1,...,5}$& 
        {$\bar{\eta'}^1$}&
        {$\bar{\eta'}^2$}&
        {$\bar{\eta'}^3$}
\\
\hline
\hline
      $\ma$ &  1 & 1&1&1 & 0,0,0,1,1 & 1 & 1 & 1 
                 & $\vec{0}$ & 0,0,0,0,0 & 0 & 0 & 0 \\
      $\ma'$&  1 & 1&1&1 & 0,0,0,0,0 & 0 & 0 & 0 
                 & $\vec{0}$ & 0,0,0,1,1 & 0 & 0 & 0 \\
\hline
\end{tabular}}
\newpage

{\bf Table B.1 Model 1 GSO Matrix $\mk$}

{\begin{tabular}{|l|rr|rrr|rrr|rr|}
\hline
 $k_{i,j}$ &$\mone$&$\mS$&$\mb_1$&$\mb_2$&$\mb_3$&$\mb'_1$&$\mb'_2$&$\mb'_3$&$\ma$&$\ma'$\\
\hline
$\mone$ & 0 & 0 & 1 & 1 & 1 & 1 & 1 & 1 & 0 & 0\\
$\mS$   & 0 & 0 & 0 & 0 & 0 & 0 & 0 & 0 & 0 & 0\\
\hline
$\mb_1$ & 1 & 1 & 1 & 1 & 1 & 0 & 0 & 0 & 0 & 1\\
$\mb_2$ & 1 & 1 & 1 & 1 & 1 & 0 & 0 & 0 & 0 & 1\\
$\mb_3$ & 1 & 1 & 1 & 1 & 1 & 0 & 0 & 0 & 0 & 1\\
\hline
$\mb'_1$& 1 & 1 & 0 & 0 & 0 & 1 & 1 & 1 & 1 & 0\\
$\mb'_2$& 1 & 1 & 0 & 0 & 0 & 1 & 1 & 1 & 1 & 0\\
$\mb'_3$& 1 & 1 & 0 & 0 & 0 & 1 & 1 & 1 & 1 & 0\\
\hline
$\ma$   & 0 & 0 & 0 & 0 & 0 & 0 & 0 & 0 & 1 & *1\\
$\ma'$  & 0 & 0 & 0 & 0 & 0 & 0 & 0 & 0 & *0& 1\\ 
\hline
\end{tabular}}
\vspace{0.4truecm}

\def\H{$\half$}

{\bf Table B.2 Model 1 Gauge Group}

{\begin{tabular}{|l|rrrrr|}
\hline
Observable &$\bar{\psi}^1$&$\bar{\psi}^2$&$\bar{\psi}^3$&$\bar{\psi}^3$&$\bar{\psi}^5$\\
\hline
\hline
$SU(4)_C$  & 0& 1&-1& 0& 0\\
           & 1& 0&-1& 0& 0\\
           & 0& 1& 1& 0& 0\\
\hline
$SU(2)_L $ & 0& 0& 0& 1&-1\\
\hline
\end{tabular}}
\vspace{0.2truecm}

{\begin{tabular}{|l|rrr|rrrrrr|rr|rrr|}
\hline 
Shadow  &$\bar{\eta}^1$&$\bar{\eta}^2$&$\bar{\eta}^3$
        &$\bar{y}_{1,3}$&$\bar{y}_{2,4}$&$\bar{y}_{5}\bar{w}_{1}$
        &$\bar{y}_{6}\bar{\omega}_{2}$&$\bar{\omega}_{3,5}$&$\bar{\omega}_{4,6}$
        &$\bar{\psi'}^4$&$\bar{\psi'}^5$
        &$\bar{\eta'}^1$&$\bar{\eta'}^2$&$\bar{\eta'}^3$\\
\hline
\hline
$SU(2)^3$
           &\H&-\H&  0&  0&\H &  0&-\H&-\H& \H& 0& 0&-\H& \H& 0\\
           &\H&  0&-\H& \H& 0 &-\H& \H&  0&-\H& 0& 0&-\H&  0& \H\\
           & 0& \H&-\H&-\H&\H & \H&  0&-\H&  0& 0& 0& 0 &-\H& \H\\
\hline
$SU(5)$  & 0&  0&  0&  0&  0&  0&  0&  0&  0& 1& 1&  0&  0& 0\\
           &\H&-\H&  0&  0&-\H&  0& \H&-\H& \H& 0& 0& \H&-\H& 0\\
           & 0& \H& \H&  0&  0&  0&-\H&  0&-\H&\H&\H&  0& \H&\H\\
           & 0& \H&-\H&-\H& \H&-\H&  0& \H&  0& 0& 0&  0& \H&-\H\\
\hline
\end{tabular}
\vspace{0.2truecm}

{\begin{tabular}{|l|rrrrr|rrrrr|}
\hline
Hidden  &$\bar{\psi}^1$&$\bar{\psi}^2$&$\bar{\psi}^3$&$\bar{\psi}^3$&$\bar{\psi}^5$
        &$\bar{\psi'}^1$&$\bar{\psi'}^2$&$\bar{\psi'}^3$&$\bar{\psi'}^3$&$\bar{\psi'}^5$\\
\hline
\hline
$SO(10)$ & 0& 0& 0& 0&  0& 0& 0& 0& 1& -1\\
           & 0& 0& 0&\H& \H& 0& 0&-1&\H&-\H\\
           & 0& 0& 0& 0&  0& 0& 1&-1& 0&  0\\
           & 0& 0& 0& 0&  0& 1&-1& 0& 0&  0\\
           & 0& 0& 0& 0&  0& 1& 1& 0& 0&  0\\
\hline
\end{tabular}}
\newpage

\def\bF{{$\bar{5}$}}
\def\bT{{$\bar{10}$}}

{\bf Table B.3 Model 1 Left-Handed States}

{\begin{tabular}{|l|rrr|rrrrrr|}
\hline
$n=1,2$&$(4_C,2_L)_O$& $(2^3,5)_{S}$&$10_{H}$&
                                            $4Q_1$&$4Q_2$&$4Q_3$&$4Q_4$&$4Q_5$&$4Q_6$\\
\hline
Singlets& &&&&&&&&\\ 
\hline
$S_{1}$      &     (1,1)&     (1,1,1,1)&     (1)&     0&  -12&   -4&  -36&  128&   80\\
$\bar{S}_{1}$&     (1,1)&     (1,1,1,1)&     (1)&     0&   12&    4&   36& -128&  -80\\
$S_{2}$      &     (1,1)&     (1,1,1,1)&     (1)&    -4&    4&    4&   -4&  176&    0\\
$\bar{S}_{2}$&     (1,1)&     (1,1,1,1)&     (1)&     4&   -4&   -4&    4& -176&    0\\
$S_{3}$      &     (1,1)&     (1,1,1,1)&     (1)&    -4&   16&    8&   32&   48&  -80\\
$\bar{S}_{3}$&     (1,1)&     (1,1,1,1)&     (1)&     4&  -16&   -8&  -32&  -48&   80\\
\hline    
Observable & &&&&&&&&\\
\hline    
$QL^{n}_{1}$&     (4,2)&     (1,1,1,1)&     (1)&     0&   -6&   -6&    6&  -40&  -80\\
$QL^{n}_{2}$&     (4,2)&     (1,1,1,1)&     (1)&     0&    0&   -4&  -56&    0&    0\\
$QL^{n}_{3}$&     (4,2)&     (1,1,1,1)&     (1)&    -6&    0&    0&    0&    0&    0\\
\hline    
$(q^{c}l^{c})^{n}_{1}$&     (-4,1)&     (2,1,1,1)&     (1)&    -2&   -4&   12&  -12&   24&  -40\\
$(q^{c}l^{c})^{n}_{2}$&     (-4,1)&     (1,2,1,1)&     (1)&     0&  -12&    8&  -28&    0&    0\\
$(q^{c}l^{c})^{n}_{3}$&     (-4,1)&     (1,1,2,1)&     (1)&     0&   -6&   10&  -10&  -64&  -40\\
\hline    
$h^{n}_{1}$&    (1,2)&   (1,1,1,1)&       (1)&     0&   -6&  -10&  -50&  -40&  -80\\
$h^{n}_{2}$&    (1,2)&   (1,1,1,1)&       (1)&    -6&   -6&   -6&    6&  -40&  -80\\
$h^{n}_{3}$&    (1,2)&   (1,1,1,1)&       (1)&    -6&    0&   -4&  -56&    0&    0\\
\hline    
$H^{n}_{1}$&    (1,2)&   (1,1,1,5)&     (1)&     2&   -8&    4&   16&   80&  -16\\
$H^{n}_{2}$&    (1,2)&   (1,1,1,5)&     (1)&     2&   10&   10&  -10&   -8&  -16\\
$H^{n}_{3}$&    (1,2)&   (1,1,1,5)&     (1)&     0&   -6&    6&   14&  -56&   64\\
\hline    
\end{tabular}}
\newpage

{\bf Table B.3 Model 1 Left-Handed States, cont.}

{\begin{tabular}{|l|rrr|rrrrrr|}
\hline
$n=1,2$&$(4_C,2_L)_O$& $(2^3,5)_{S}$&$10_{H}$&
                                            $4Q_1$&$4Q_2$&$4Q_3$&$4Q_4$&$4Q_5$&$4Q_6$\\
\hline
Shadow      & &&&&&&&&\\
\hline
$F_{1}$&      (1,1)&   (1,1,1,5)&       (1)&     0&  -12&   -4&  -36&  -96&  -16\\
$\bar{F}_{1}$&(1,1)&   (1,1,1,\bF)&     (1)&     0&   12&    4&   36&   96&   16\\
$F_{2}$&      (1,1)&   (1,1,1,5)&       (1)&    -4&   -8&    0&  -40&   80&  -16\\
$\bar{F}_{2}$&(1,1)&   (1,1,1,\bF)&     (1)&     4&    8&    0&   40&  -80&   16\\
$F_{3}$&      (1,1)&   (1,1,1,5)&       (1)&    -4&    4&    4&   -4&  -48&  -96\\
$\bar{F}_{3}$&(1,1)&   (1,1,1,\bF)&     (1)&     4&   -4&   -4&    4&   48&   96\\
\hline    
$\bar{G}^{n}_{1}$&(1,1)& (2,1,1,\bT)&     (1)&     0&   -6&   -2&  -18&  -48&   -8\\
$\bar{G}^{n}_{2}$&(1,1)& (1,2,1,\bT)&     (1)&    -2&    2&    2&   -2&  -24&  -48\\
$\bar{G}^{n}_{3}$&(1,1)& (1,1,2,\bT)&     (1)&    -2&   -4&    0&  -20&   40&   -8\\
\hline 
$X^{n}_{1}$      &(1,1)& (2,1,1,1)&       (1)&    -2&   -4&   16&   44&   24&  -40\\
$X^{n}_{2}$      &(1,1)& (2,1,1,1)&       (1)&    -2&    2&   18&  -18&   64&   40\\
$X^{n}_{3}$      &(1,1)& (2,1,1,1)&       (1)&    -2&    2&  -14&   14&  112&  -40\\
$X^{n}_{4}$      &(1,1)& (2,1,1,1)&       (1)&    -2&   20&   -8&  -12&   24&  -40\\
\hline 
$Y^{n}_{1}$      &(1,1)& (1,2,1,1)&       (1)&     0&   -6&  -18&   -2&   88&    0\\
$Y^{n}_{2}$      &(1,1)& (1,2,1,1)&       (1)&     6&  -12&    8&  -28&    0&    0\\
$Y^{n}_{3}$      &(1,1)& (1,2,1,1)&       (1)&     0&   -6&   14&  -34&   40&   80\\
$Y^{n}_{4}$      &(1,1)& (1,2,1,1)&       (1)&    -2&   -4&  -16&   -4&  -48&   80\\
\hline 
$Z^{n}_{1}$      &(1,1)& (1,1,2,1)&       (1)&    -2&    2&  -14&   14& -112&   40\\
$Z^{n}_{2}$      &(1,1)& (1,1,2,1)&       (1)&     0&   -6&   14&   46&  -64&  -40\\
$Z^{n}_{3}$      &(1,1)& (1,1,2,1)&       (1)&     0&   18&  -10&  -10&  -64&  -40\\
$Z^{n}_{4}$      &(1,1)& (1,1,2,1)&       (1)&     6&   -6&   10&  -10&  -64&  -40\\
\hline 
$U_{1}$          &(1,1)& (2,2,1,1)&       (1)&    -4&   -8&    0&   40&  -24&   40\\
$\bar{U}_{1}$    &(1,1)& (2,2,1,1)&       (1)&     4&    8&    0&  -40&   24&  -40\\
$U_{2}$          &(1,1)& (2,1,2,1)&       (1)&    -4&    4&    4&   -4&  -48&   80\\
$\bar{U}_{2}$    &(1,1)& (2,1,2,1)&       (1)&     4&   -4&   -4&    4&   48&  -80\\
$U_{3}$          &(1,1)& (1,2,2,1)&       (1)&     0&  -12&   -4&   44&   24&  -40\\
$\bar{U}_{3}$    &(1,1)& (1,2,2,1)&       (1)&     0&   12&    4&  -44&  -24&   40\\
\hline 
Hidden& &&&&&&&&\\
\hline
$T^{n}_{1}$      &(1,1)& (1,1,1,1)&      (16)&    -2&  -16&    0&    0&    0&    0\\
$T^{n}_{2}$      &(1,1)& (1,1,1,1)&      (16)&    -2&    2&    6&  -26&  -88&    0\\
$T^{n}_{3}$      &(1,1)& (1,1,1,1)&      (16)&     0&    0&    4&  -24&  -24&    0\\
\hline    
\end{tabular}}
\newpage

{\bf Table C.1 Model 2 GSO Matrix $\mk$}

{\begin{tabular}{|l|rr|rrr|rrr|rr|}
\hline
 $k_{i,j}$ &$\mone$&$\mS$&$\mb_1$&$\mb_2$&$\mb_3$&$\mb'_1$&$\mb'_2$&$\mb'_3$&$\ma$&$\ma'$\\
\hline
$\mone$ & 0 & 0 & 1 & 1 & 1 & 1 & 1 & 1 & 0 & 0\\
$\mS$   & 0 & 0 & 0 & 0 & 0 & 0 & 0 & 0 & 0 & 0\\
\hline
$\mb_1$ & 1 & 1 & 1 & 1 & 1 & 0 & 0 & 0 & 0 & 0\\
$\mb_2$ & 1 & 1 & 1 & 1 & 1 & 0 & 0 & 0 & 0 & 0\\
$\mb_3$ & 1 & 1 & 1 & 1 & 1 & 0 & 0 & 0 & 0 & 0\\
\hline
$\mb'_1$& 1 & 1 & 0 & 0 & 0 & 1 & 1 & 1 & 0 & 0\\
$\mb'_2$& 1 & 1 & 0 & 0 & 0 & 1 & 1 & 1 & 0 & 0\\
$\mb'_3$& 1 & 1 & 0 & 0 & 0 & 1 & 1 & 1 & 0 & 0\\
\hline
$\ma$   & 0 & 0 & 1 & 1 & 1 & 0 & 0 & 0 & 1 & *1\\
$\ma'$  & 0 & 0 & 0 & 0 & 0 & 1 & 1 & 1 & *0 & 1\\ 
\hline
\end{tabular}}
\vspace{0.4truecm}

{\bf Table C.2 Model 2 Gauge Group}

{\begin{tabular}{|l|rrrrr|}
\hline
Observable &$\bar{\psi}^1$&$\bar{\psi}^2$&$\bar{\psi}^3$&$\bar{\psi}^3$&$\bar{\psi}^5$\\
\hline
\hline
$SU(4)_C$  & 0& 1&-1& 0& 0\\
           & 1& 0&-1& 0& 0\\
           & 0& 1& 1& 0& 0\\
\hline
$SU(2)_R $ & 0& 0& 0& 1& 1\\
\hline
\end{tabular}}
\vspace{0.2truecm}

{\begin{tabular}{|l|rrr|rrrrrr|rrr|}
\hline
Shadow  &$\bar{\eta}^1$&$\bar{\eta}^2$&$\bar{\eta}^3$
        &$\bar{y}_{1,3}$&$\bar{y}_{2,4}$&$\bar{y}_{5}\bar{w}_{1}$
        &$\bar{y}_{6}\bar{\omega}_{2}$&$\bar{\omega}_{3,5}$&$\bar{\omega}_{4,6}$
        &$\bar{\eta'}^1$&$\bar{\eta'}^2$&$\bar{\eta'}^3$\\
\hline
\hline
$SU(2)^3$
           & $\half$&$-\half$& 0& 0&$\half$& 0&$-\half$&$-\half$&$\half$&$-\half$&$\half$& 0\\
           & $\half$& 0&$-\half$&$\half$& 0&$-\half$&$\half$& 0&$-\half$&$-\half$& 0& $\half$\\
           & 0&$\half$&$-\half$&$-\half$&$\half$&$\half$& 0&$-\half$& 0 & 0&$-\half$& $\half$\\
\hline
$SU(3)$  &$\half$&$-\half$& 0& 0&$-\half$& 0&$\half$&$-\half$&$\half$ &$\half$&$-\half$& 0\\
           &$\half$& 0&$-\half$&$-\half$& 0&$-\half$&$\half$& 0&$\half$ &$\half$& 0&$-\half$\\
\hline
\end{tabular}}
\vspace{0.2truecm}

{\begin{tabular}{|l|rrrrr|rrrrr|}
\hline
Hidden  &$\bar{\psi}^1$&$\bar{\psi}^2$&$\bar{\psi}^3$&$\bar{\psi}^3$&$\bar{\psi}^5$
        &$\bar{\psi'}^1$&$\bar{\psi'}^2$&$\bar{\psi'}^3$&$\bar{\psi'}^3$&$\bar{\psi'}^5$\\
\hline
\hline
$SO(10)$    & 0& 0& 0& 0& 0            & 0& 0& 0& 1& 1\\
              & 0& 0& 0&$\half$&$-\half$ & 0& 0&-1&$\half$&$\half$\\
              & 0& 0& 0& 0& 0            & 0& 1&-1& 0& 0\\
              & 0& 0& 0& 0& 0            & 1&-1& 0& 0& 0\\
              & 0& 0& 0& 0& 0            & 1& 1& 0& 0& 0\\
\hline
$SU(2)_L$ & 0& 0& 0& 0& 0            & 0& 0& 0& 1&-1\\
\hline
\end{tabular}}
\newpage

\def\bt{{$\bar{3}$}}
     
{\bf Table C.3 Model 2 Left-Handed States}

{\begin{tabular}{|l|rrr|rrrrrrr|}
\hline
$n=1,2$ & $(4_C,2_R)_O$& $(2^3,3)^S$&$(10,2_R)^H$&
                                            $4Q_1$&$4Q_2$&$4Q_3$&$4Q_4$&$4Q_5$&$4Q_6$&$4Q_7$\\
\hline
Singlets& &&&&&&&&&\\ 
\hline
 $S^{n}_{1}$&     (1,1)&     (1,1,1,1)&     (1,1)&    12&    0&   -4&   -8&   16&    0&    0\\
 $S^{n}_{2}$&     (1,1)&     (1,1,1,1)&     (1,1)&    12&    0&   -4&    8&  -16&    0&    0\\
 $S^{n}_{3}$&     (1,1)&     (1,1,1,1)&     (1,1)&    12&    0&    0&   -6&    0&    4&   24\\
 $S^{n}_{4}$&     (1,1)&     (1,1,1,1)&     (1,1)&    12&    0&    0&    6&    0&    4&   24\\
 $S^{n}_{5}$&     (1,1)&     (1,1,1,1)&     (1,1)&    12&    0&    0&    2&  -16&    4&  -24\\
 $S^{n}_{6}$&     (1,1)&     (1,1,1,1)&     (1,1)&    12&    0&    0&   -2&   16&    4&  -24\\
 $S_{7}$&         (1,1)&     (1,1,1,1)&     (1,1)&     0&    0&    0&    8&   32&    0&    0\\
 $\bar{S}_{7}$&   (1,1)&     (1,1,1,1)&     (1,1)&     0&    0&    0&   -8&  -32&    0&    0\\
 $S_{8}$&         (1,1)&     (1,1,1,1)&     (1,1)&     0&    0&    8&    4&   16&    8&    0\\
 $\bar{S}_{8}$&   (1,1)&     (1,1,1,1)&     (1,1)&     0&    0&   -8&   -4&  -16&   -8&    0\\
 $S_{9}$&         (1,1)&     (1,1,1,1)&     (1,1)&     0&    0&    8&   -4&  -16&    8&    0\\
 $\bar{S}_{9}$&   (1,1)&     (1,1,1,1)&     (1,1)&     0&    0&   -8&    4&   16&   -8&    0\\
\hline    
Observable&&&&&&&&&&\\
\hline
$QL^{n}_{1}$&     (4,2)&     (1,1,1,1)&     (1,1)&    -6&   -2&   -2&    2&  -16&    0&   -8\\
$QL^{n}_{2}$&     (4,2)&     (1,1,1,1)&     (1,1)&    -6&   -2&   -2&   -2&   16&    0&   -8\\
$QL^{n}_{3}$&     (4,2)&     (1,1,1,1)&     (1,1)&    -6&   -2&    2&    0&    0&    4&   16\\
\hline    
$(q^{c}l^{c})^{n}_{1}$&    (-4,1)&     (2,1,1,1)&     (1,1)&     6&   -6&   -2&    0&    0&   -4&    0\\
$(q^{c}l^{c})^{n}_{2}$&    (-4,1)&     (1,2,1,1)&     (1,1)&     6&   -6&    2&   -2&   -8&    0&    0\\
$(q^{c}l^{c})^{n}_{3}$&    (-4,1)&     (1,1,2,1)&     (1,1)&     6&   -6&    2&    2&    8&    0&    0\\
\hline    
$h_{1}$&         (1,2)&   (2,1,1,1)&       (1,1)&     0&   -8&    0&    0&    0&    0&   16\\
$\bar{h}_{1}$&   (1,2)&   (2,1,1,1)&       (1,1)&     0&    8&    0&    0&    0&    0&  -16\\
$h_{2}$&         (1,2)&   (1,2,1,1)&       (1,1)&     0&   -8&    0&   -4&    8&    0&   -8\\
$\bar{h}_{2}$&   (1,2)&   (1,2,1,1)&       (1,1)&     0&    8&    0&    4&   -8&    0&    8\\
$h_{3}$&         (1,2)&   (1,1,2,1)&       (1,1)&     0&   -8&    0&    4&   -8&    0&   -8\\
$\bar{h}_{3}$&   (1,2)&   (1,1,2,1)&       (1,1)&     0&    8&    0&   -4&    8&    0&    8\\
\hline    
$H^{n}_{1}$&     (1,2)&   (2,1,1,3)&       (1,1)&     4&    0&    4&    0&    0&    0&    0\\
$H^{n}_{2}$&     (1,2)&   (1,2,1,3)&       (1,1)&     4&    0&    0&    2&    8&   -4&    0\\
$H^{n}_{3}$&     (1,2)&   (1,1,2,3)&       (1,1)&     4&    0&    0&   -2&   -8&   -4&    0\\
\hline    
$B^{n}_{1}$&     (6,1)&   (1,1,1,1)&       (1,1)&     0&    4&    4&    6&    0&   -4&   -8\\
$B^{n}_{2}$&     (6,1)&   (1,1,1,1)&       (1,1)&     0&    4&    0&    0&    0&   -8&   16\\
$B^{n}_{3}$&     (6,1)&   (1,1,1,1)&       (1,1)&     0&    4&    4&   -6&    0&   -4&   -8\\
\hline
\end{tabular}}
\newpage

{\bf Table C.3 Model 2 Left-Handed States cont.}

{\begin{tabular}{|l|rrr|rrrrrrr|}
\hline
$n=1,2$ & $(4_C,2_R)_O$& $(2^3,3)^S$&$(10,2_R)^H$&
                                            $4Q_1$&$4Q_2$&$4Q_3$&$4Q_4$&$4Q_5$&$4Q_6$&$4Q_7$\\
\hline
Shadow & &&&&&&&&&\\
\hline
$F^{n}_{1}$&     (1,1)&   (1,1,1,\bt)&       (1,1)&    -4&   -8&   -4&    0&    0&    0&   16\\
$F^{n}_{2}$&     (1,1)&   (1,1,1,\bt)&       (1,1)&    -4&   -8&    0&   -6&    0&    4&   -8\\
$F^{n}_{3}$&     (1,1)&   (1,1,1,\bt)&       (1,1)&    -4&   -8&    0&    6&    0&    4&   -8\\
$F^{n}_{4}$&     (1,1)&   (1,1,1,\bt)&       (1,1)&    -4&    8&   -4&    0&    0&    0&  -16\\
$F^{n}_{5}$&     (1,1)&   (1,1,1,\bt)&       (1,1)&    -4&    8&    0&   -2&   16&    4&    8\\
$F^{n}_{6}$&     (1,1)&   (1,1,1,\bt)&       (1,1)&    -4&    8&    0&    2&  -16&    4&    8\\
\hline    
$F^{n}_{7}$&     (1,1)&   (1,1,1,3)  &       (1,1)&    -8&    0&   -8&    0&    0&    0&    0\\
$\bar{F}^{n}_{7}$&(1,1)&  (1,1,1,\bt)&       (1,1)&     8&    0&    8&    0&    0&    0&    0\\
$F^{n}_{8}$&     (1,1)&   (1,1,1,3)  &       (1,1)&    -8&    0&    0&   -4&  -16&    8&    0\\
$\bar{F}^{n}_{8}$&(1,1)&  (1,1,1,\bt)&       (1,1)&     8&    0&    0&    4&   16&   -8&    0\\
$F^{n}_{9}$&     (1,1)&   (1,1,1,3)  &       (1,1)&    -8&    0&    0&    4&   16&    8&    0\\
$\bar{F}^{n}_{9}$&(1,1)&  (1,1,1,\bt)&       (1,1)&     8&    0&    0&   -4&  -16&   -8&    0\\
\hline 
$X^{n}_{1}$&     (1,1)&   (2,2,1,1)&       (1,1)&   -12&    0&    0&   -2&   -8&   -4&    0\\
$X_{2}$&         (1,1)&   (2,2,1,1)&       (1,1)&     0&    0&    0&   -4&    8&    0&  -24\\
$X_{3}$&         (1,1)&   (2,2,1,1)&       (1,1)&     0&    0&    0&    4&   -8&    0&   24\\
\hline    
$Y^{n}_{1}$&     (1,1)&   (2,1,2,1)&       (1,1)&   -12&    0&    0&    2&    8&   -4&    0\\
$Y_{2}$&         (1,1)&   (2,1,2,1)&       (1,1)&     0&    0&    0&    4&   -8&    0&  -24\\
$Y_{3}$&         (1,1)&   (2,1,2,1)&       (1,1)&     0&    0&    0&   -4&    8&    0&   24\\
\hline    
$Z^{n}_{1}$&     (1,1)&   (1,2,2,1)&       (1,1)&   -12&    0&    4&    0&    0&    0&    0\\
$Z_{2}$&         (1,1)&   (1,2,2,1)&       (1,1)&     0&    0&    0&   -8&   16&    0&    0\\
$Z_{3}$&         (1,1)&   (1,2,2,1)&       (1,1)&     0&    0&    0&    8&  -16&    0&    0\\
\hline 
Hidden& &&&&&&&&&\\
\hline
$T^{n}_{1}$&      (1,1)&   (1,1,1,1)&      (10,1)&    0&   -4&    4&   -2&   16&   -4&    8\\
$T^{n}_{2}$&      (1,1)&   (1,1,1,1)&      (10,1)&    0&   -4&    4&    2&  -16&   -4&    8\\
$T^{n}_{3}$&      (1,1)&   (1,1,1,1)&      (10,1)&    0&   -4&    0&    0&    0&   -8&  -16\\
\hline    
$\bar{\it F}^{n}_{1}$&     (1,1)&   (1,1,1,\bt)&  (1,2)&   -4&   -4&    4&    0&    0&    0&  -16\\
$\bar{\it F}^{n}_{2}$&     (1,1)&   (1,1,1,\bt)&  (1,2)&   -4&   -4&    0&   -6&    0&   -4&    8\\
$\bar{\it F}^{n}_{3}$&     (1,1)&   (1,1,1,\bt)&  (1,2)&   -4&   -4&    0&    6&    0&   -4&    8\\
\hline    
$D^{n}_{1}$&     (1,1)&   (1,1,1,1)&       (1,2)&    12&    4&    4&    0&    0&    0&   16\\
$D^{n}_{2}$&     (1,1)&   (1,1,1,1)&       (1,2)&    12&    4&    0&   -2&   16&   -4&   -8\\
$D^{n}_{3}$&     (1,1)&   (1,1,1,1)&       (1,2)&    12&    4&    0&    2&  -16&   -4&   -8\\
\hline    
\end{tabular}}
\newpage

{\bf Table D.1 Hypercharge Components.}

{\begin{tabular}{|l|rrr|cl|}
\hline
   & $\tilde{Q}_{B-L}$& $2 T_3^R$    & $2 T_3^L$   & $\tilde{Q}_{EM}$ &$= T_3^L + T_3^R + \half \tilde{Q}_{B-L}$ \\
   & $\third \tilde{Q}_C$ & $\half \tilde{Q}_L $ &             &          & $= T_3^L + \half Y $ \\
\hline
\hline
$Q_L$        & $ \third$   &$  0$  & $\pm 1$     & &$\twothird$, $-\third$  \\ 
$d^{c}_L$  & $-\third$   &$  1$  & $    0$     & &$\third$                \\ 
$u^{c}_L$  & $-\third$   &$ -1$  & $    0$     & &$-\twothird$            \\ 
$L_L$        &  $-1    $   &$  0$  & $\pm 1$     & &$0$, $-1$               \\ 
$e^{c}_L$  &  $ 1    $   &$  1$  & $    0$     & &$1$                     \\ 
$\nu^{c}_L$&  $ 1    $   &$ -1$  & $    0$     & &$0$                     \\ 
\hline    
\end{tabular}}
\end{document}